# On the Limits of Perfect Security for Steganographic System

Khan Farhan Rafat, M. Sher

Department of Computer Science, International Islamic University
Islamabad, 44000, Pakistan

**Abstract**
Until now the discussion on perfect security for steganographic systems has remained confined within the realm of mathematicians and information theory experts whose concise and symbolic representation of their philosophies, postulates, and inference thereafter has made it hard for the naïve academics to have an insight of the concepts. This paper is an endeavor not only to appraise on the limitations of one of such pioneer comprehensions but also to illustrate a pitfall in another scheme that asserts on having perfect security without the use of public or secret key. Goals set are accomplished through contrasting test results of a steganographic scheme that exploits English words with corresponding acronyms for hiding bits of secret information in chat - a preferred way to exchange messages these days. The misapprehension about perfect security and reign in characteristic of stego key in bit embedding process are unfolded respectively by launching elementary chosen-message and chosen-cover attack, and through proposed enhancement of target scheme.
**Keywords:** *Perfect Security, Conceal, Covert Channel, Deception, Oblivious Communication, Unobtrusive.*

## 1. Introduction

*Organized data* serves as a single point of reference that makes prediction about an event a much simpler task than without it i.e., provisioning of some sort of meaningful *information* for decision making, the essence of which, however, cannot be measured but the severity of which can be realized from the fact that in today's world information is being regarded as a *double-edged sword* [1] capable of imparting devastating impact on the privacy of an individual and as well as nation (at global level) more withering than what was faced by Hiroshima and Nagasaki during World War - II.

Some of the most recent incidents including *Blaster worm* attack (2003), electronic cutoff of Estonia from rest of the world (2007), attack by *StuXnet* virus on Iranian nuclear installations (2010) and Wiki leaks (2011) calls for a daring need, more than ever before, to guard information security frontiers by evolving new and analyzing and modifying/updating existing security schemes from falling into the hands of hostile.

Steganography is referred to as art and science [2] for covert communication. The name was first cited in a work presented by *Trithemus* (1462-1516) entitled *Steganographia* and is of Greek origin where the words στεγανό-ς (Steganos) and γραφ-ειν (Graphos) are put together as single English word which means covered/Hidden Writing [3]. The essence of steganography is to hide the very existence of information [4] in contrast to cryptography whose rationale is to make information incomprehensible [5]. By virtue of being seamless steganography has emerged as a preferred choice for information hiding these days.

1.1 Paper Plot

This paper is organized as follows: Section 2 expound on the limitations of perfect security and consequence of its realization without using stego key. Section 3 elaborates on bit embedding scheme with the help of which short comings of prevalent apprehensions will be emphasized. Section 4 highlights on tools used to contrast resemblance/dissimilarity between cover text and stego object. Analysis of the target scheme using test cases along with test results are shown in Section 5. Reign in attribute of stego key towards system's security is discussed in Section 6 where an enhancement in context of information theoretic security is proposed. Section 7 concludes on our argument.

## 2. Perfect Security

Subsequent discussion expands on the notion of perfect security and allied misapprehensions. However, as already stated, deliberate effort has been made to forgo complex mathematical illustrations, elaborate and exemplify the concept in simple terms for easy understanding.

2.1 Cachin's Conception

Cachin [6] was first to come up with the idea of perfect security by suggesting for an information theoretically secure scheme where seamless bit embedding in randomly selected cover text renders a stego object that has the same probability distribution as that of the former and equated it as follows:

$$D(P_c \| P_s) \leftarrow \varepsilon \leftarrow \sum_{q \in Q} P_c(q) \log_2 \frac{P_c(q)}{P_s(q)} \quad (1)$$

, where $P_c$ & $P_s$ are probability distributions for cover text and stego object respectively.
It is obvious that $\frac{P_c(q)}{P_s(q)} = 0$ for $\frac{0}{P_s(q)}$, and equals $\infty$ for $\frac{P_c(q)}{0}$.

### 2.1.1 Precincts

i. Evidently for $\varepsilon = 0$, the system is perfectly secure. But the question is whether such system exist? According to Cachin such a system does exist -at least theoretically- called *One Time Pad* (OTP) but realized that how *Wendy* (an observer) would let go such an output (stego object).
ii. It is apparent from equation (1) that bit embedding has an implicit binding to render an output that must be from within the random oracle constituting the cover texts or else $\varepsilon \to \infty$ i.e. alphabet's bound constraint.
iii. Notwithstanding above the trait to undermine Wendy's ability to detect such communication seems fictitious in its eternity.

### 2.2 Perfect Security without Public or Secret Key

Boris Ryabko and Daniil Ryabko [7] opted for generating all possible fixed length sequences of a given cover text and then transmitting only the sequence number that corresponds to the secret bits of the message where the block length of cover text must also be ≤ message bit stream.

### 2.2.1 Limitations

i. Besides having obvious memory constraint the provision of '*given cover text*' at both ends covertly accentuate on a shared secret (i.e. *key*) without which it cannot be regarded as an oblivious communication (*A contradiction of the postulate in itself*).
ii. Knowledge of the algorithm alone is sufficient to retrieve hidden sequence number to exact on the message bits.
iii. Contradicts Kerchoff's Principle [8].

## 3. Target Steganographic Scheme

Table 1 – List of Words along with their Acronyms

| Acronyms Column Label (1) | Words/Phrases Column Label (0) |
|---|---|
| & | and |
| 2 | To |
| 2NIGHT | Tonight |
| 4 | For |
| … | … |
| 4U | For you |
| … | … |
| A3 | Anytime, anywhere, anyplace |
| AAMOF | As a matter of fact |
| ABH | Anyone but him |
| ACTO | According to |
| … | … |
| ADD | Address |
| ADR | Ain't doin' right |
| AE | Almost every |
| AFAIC | As far as I'm concerned |
| … | … |

To precisely apprehend on our revelation we selected the chat-lingo of the so called 'thumb generation' which is easy to grasp and does not arouse suspicion. Acronyms are contractions for comparatively long or frequently used words/phrases like *As soon as possible* which is abbreviated as *ASAP*. [9] Suggested using acronyms together with corresponding words / phrases of English language to hide bits of secret information in chat. The scheme works by arranging words/phrases in one of the two column table, the other column of which is populated with corresponding acronyms. The column containing words/phrases are labeled as "0" while acronyms are headed by label "1". Table 1 indicates one such arrangement.

Next, text cover composed of words/phrases and acronym from the predefined table is prepared. Secret information to be hidden inside the body of text cover is translated into bits. Text cover is then iterated till its end to search for words/phrase or acronym matching those in the table till end. Each time when a word/phrase or acronym is encountered corresponding secret message bit (in sequence) is examined with reference to column heads (label) of the table. Suppose if secret message bit is 0 and the corresponding matching text in cover is word/phrase i.e., having column labeled as 0, the cover text remains unchanged. However, if the secret message bit is 1 and the corresponding matching text in cover is word/phrase, the word/phrase in cover text is replaced by its corresponding acronym. In short, binary message bit 0 corresponds to having word/phrase in the stego object while binary message bit 1 corresponds to having acronym in place of words/phrases.

## 4. Tools to Realize Quality of Test Results

To maintain transparency and for better understanding and visual substantiation of test results we favored for probability distribution plots between cover text and stego object using Minitab 16 [10] and presented our reader with quantified output through Hamming [11], Levenshtein (Edit Distance) [12], and Jaro-Winkler [13] distance.

## 5. Analysis

To analyze our target scheme we arranged English Language words and their corresponding acronyms derived from [14] in two columns and labeled those as (0) and (1) respectively as in Table 1.

Next was the choice of secret message to unveil limitations of the idea of perfect security. Since, the target steganographic scheme allows for embedding of secret message bits (0/1) inside text-cover, hence, we favored for testing the scheme using message bits which were comprised of all 0's and developed three test cases as under:

***Case – I:*** *Cover text comprising of regular text but without acronyms (i.e. cover text devoid of acronyms):*

It is evident from Table 1 that for text cover comprising of all words *(exclusive of words having acronyms)*, a chosen-message of all 0-bits resulted in stego object which is an exact replica of the cover text.

a. *Quantified Test Results*

A. Hamming, Levenshtein, and Jaro-Winkler distance for cover text and stego Object are computed as shown in Table 2.

Table 2 – Quantifying closeness of Cover Text & Stego Object

| Distance | Difference |
|---|---|
| Hamming | 0 |
| Levenshtein | 0 |
| Jaro-Winkler | 0 |

B. To contrast probability distribution plots (Fig. 1 refers) of cover text and stego object, their mean, variance and standard deviation are computed as shown in Table 3.

Table 3 – Quantified Similarity between Cover Text & Stego Object

|  | Cover Text | Stego Object | Difference |
|---|---|---|---|
| Mean | 0.089 | 0.089 | 0 |
| Variance | 0.976 | 0.976 | 0 |
| STD | 0.031 | 0.031 | 0 |

b. *Envisaged Test Results*

Figure 1 illustrates an exact match between cover text and stego object that going with the notion of perfect security renders $\varepsilon = 0$ i.e. appears as **PERFECTLY SECURE**. We, however, envisage the test result as frightening as **$\varepsilon = 0$ does not account for 'perfect security'**.

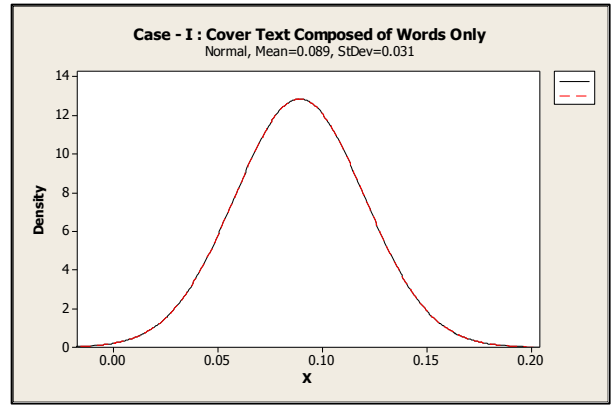

Fig. 1 Probability Distribution Graphs Contrasting Cover Text & Stego Object

**Case – II:** *Cover text comprising of regular text and all acronyms in place of their corresponding words:*

We repeated the preceding course and test results obtained thereafter are elaborated as under:

a. *Quantified Test Results.*

A. Hamming, Levenshtein, and Jaro-Winkler distance for cover text and stego (Text) Object are computed as shown in Table 4.

Table 4 – Quantifying closeness of Cover Text & Stego Object

| Distance | Difference |
|---|---|
| Hamming | Different file lengths |
| Levenshtein | 228 |
| Jaro-Winkler | 0.443 |

B. To contrast probability distribution plots (Fig. 2 refers) of cover text and stego object, their mean, variance and standard deviation are computed as shown in Table 5.

Table 5 – Quantified Similarity between Cover Text & Stego Object

|          | Cover Text | Stego Object | Difference |
|----------|------------|--------------|------------|
| **Mean** | 0.058 | 0.089 | 0.031 |
| **Variance** | 0.315 | 0.976 | 0.661 |
| **STD** | 0.017 | 0.031 | 0.014 |

b. *Envisaged Test Results*

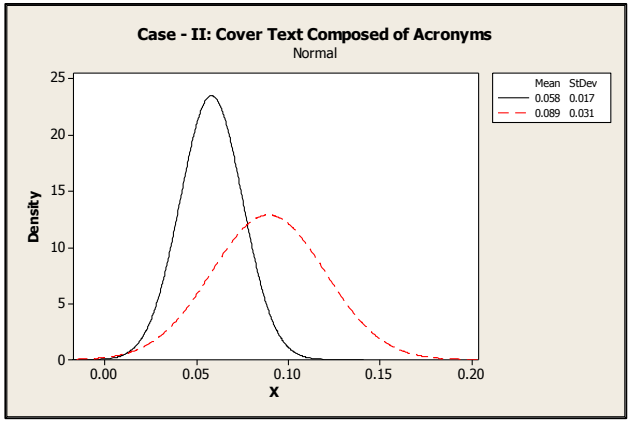

Fig. 2  Probability Distribution Graphs Contrasting Cover Text & Stego Object

The similarity between cover and stego object can be realized from Fig. 2 where the two probability distribution graphs vary for their mean and standard deviation by values 0.031 and 0.014 respectively. However, *interestingly all acronyms in cover text were replaced by their corresponding word/phrase*.

But is this scheme secure? We, however, doubt it since **a chosen message alone is sufficient to expose bit embedding methodology i.e. words having acronyms conceal secret binary message bit '0'.**

Case – III: *Cover text comprising of regular English text and mix of words with their corresponding acronyms*

Likewise, we experimented by composing cover text with a mix of words along with their corresponding acronyms whose test results follow as under:

a. *Quantified Test Results*

A. Hamming, Levenshtein, and Jaro-Winkler distance for cover text and stego (Text) Object are computed as shown in Table 6.

Table 6 – Quantifying closeness of Cover Text & Stego Object

| Distance | Difference |
|----------|------------|
| Hamming | Different file lengths |
| Levenshtein | 326 |
| Jaro-Winkler | 0.902 |

B. To contrast probability distribution plots (Fig. 3 refers) of cover text and stego object, their mean, variance and standard deviation are computed as shown in Table 7.

Table 7 – Quantified Similarity between Cover Text & Stego Object

|          | Cover Text | Stego Object | Difference |
|----------|------------|--------------|------------|
| **Mean** | 0.082 | 0.080 | 0.031 |
| **Variance** | 0.988 | 0.955 | 0.661 |
| **STD** | 0.031 | 0.030 | 0.014 |

b. *Envisaged Test Results*

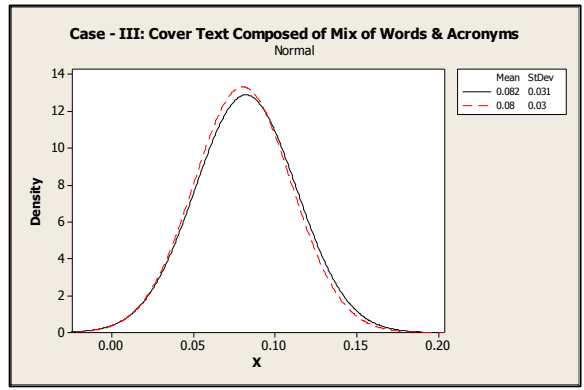

Fig. 3  Probability Distribution Graphs Contrasting Cover Text & Stego Object

The similarity between cover and stego object can be realized from Fig. 3 where the probability distribution graphs vary for their mean and standard deviation by values 0.031 and 0.014 respectively. Attention, however, is invited to the fact that *here words and acronyms in cover text were replaced by their corresponding acronyms and words in stego object respectively*.

**Above, however, once again reflects on the limitation of the existing scheme and reconfirms on the deficiency of some shared secret (i.e. stego key etc.) between communicating parties**.

Similarly, repeating the aforesaid cases with a chosen-message comprising of all binary bits '1', and subsequently by contrasting differences in known-cover and resultant stego object, reconfirmed our findings.

## 6. Proposed Enhancement

The trait of aforementioned scheme also fell well short of Kerckhoff's principle as knowledge of algorithm or cover or experimenting with choice of message alone suffices to unveil bit embedding methodology.

Based on its logical formation we opted for information theoretic model [15] where pre-processing of covert text just before bit embedding is proposed and ensued as follows:

6.1 Evolved Algorithm

Prepared cover text comprising of English words including acronyms. Generated 256-bit stego key (FIPS publication 198) [16]. Obtained 256-bit HASH of the stego key using SHAH-256 [17]. We then iterated through the entire cover text once. Searching for words or acronyms in Table 1 and swapping those with their corresponding acronyms or words respectively in the cover text if corresponding binary bit of the HASH is '1' i.e. *random pre-processing of the cover text*.

Next we iterated and swapped column entries of Table 1 for corresponding binary bit '1' of the stego key and obtained a new Table 8 as in [18].

Table 8 – Mixed List of Words and their Corresponding Acronyms

| **Acronyms** **Column Label (1)** | **Stego Key** | **Words/Phrases Column Label (0)** |
|---|---|---|
| and | 1 | & |
| 2 | 0 | To |
| Tonight | 1 | 2NIGHT |
| For | 1 | 4 |
| … | … | … |
| 4U | 0 | For you |
| … | … | … |
| Anytime, anywhere, anyplace | 1 | A3 |
| As a matter of fact | 1 | AAMOF |
| ABH | 0 | Anyone but him |
| ACTO | 0 | According to |
| … | … | … |
| ADD | 0 | Address |
| ADR | 0 | Ain't doin' right |
| Almost every | 1 | AE |
| As far as I'm concerned | 1 | AFAIC |
| … | … | … |

Since entries in Table 1 are far more than 256, hence the pointer to stego key bits is re initialized to 1 each time it exceeds the value 256 i.e. *random shuffling of the contents of pre-agreed list*.

On the analogy of Section 5, we composed (chose) a secret message encompassing all '0'-bits and proceeded with the three test cases respectively by iterating through the preprocessed cover text and substituting words or acronyms with words/acronyms from column labeled as '0' of Table 8 – corresponding to chosen-message bit '0'.

6.2 Test Results

Interestingly the outcomes of the three test cases have almost proportionate distribution of words and acronyms for the same message which were spread over the entire stego object, similar to those envisage in Table 7, thereby affixing on the suitability of our proposed enhancement even for worst case scenarios and assuring its decoding only by those in possession of stego key (adherence to Kerchoff's principle) or via brute force attack. . *The preprocessing stage introduces an uncertainty that is difficult to comprehend.*

6.3 Future Work

Following in sequence may further add operational ease along with entailing Wendy's efforts to extract the hidden secret out of the cover text:

i. Type of message i.e. text, image etc. and message length be appended before secret message as header.
ii. Stego key dependent bisection of the message along with header.
iii. Compression of bisected secret message.
iv. Encryption of compressed data before bit embedding.
v. Adding digital signatures will ensure message integrity and non-repudiation.
vi. Use of PKI etc.to exchange stego keys.

## 7. Conclusion

**Security can neither be measured in terms of length and breadth nor can it be quantified.** *The notion on perfect security for finite sequence and without a shared secret between communicating parties seems fictitious* as no matter what, a finite sequence will always get decoded constraint only by time, resources and determination of the adversary e.g. by launching brute force attack. Hence the safest and most pragmatic approach is not to underestimate Wendy's ability to precise on cover text carrying hidden message but rather randomness be made a part of stego key dependent bit embedding process (*Stego key* be preferably unique in its perpetuity) to have varied outcomes even for the same message and using same cover text. It is only through this postulation that we are ought to devise information theoretically secure schemes that can prolong Wendy's efforts for a time equal to or longer than that where the hidden information losses its vitality.

Foresaid above, ours is also the first known attempt to scrutinize the cogency of 'perfect security' by launching basic chosen-message and chosen-cover attacks on an unpretentious steganographic scheme.

## References

[1] Kendall Hoyt, Stephen g. Brooks, A Double-Edged Sword: Globalization and Biosecurity. International Security, Vol. 28, No. 3, pp. 123-148, Winter 2003/04.

[2] Chincholkar A.A. and Urkude D.A. (2012) Design and Implementation of Image Steganography. Journal of Signal and Image Processing, ISSN: 0976-8882 & E-ISSN: 0976-8890, Volume 3, Issue 3, pp.-111-113.

[3] Alexander A. Sherstov. The Communication Complexity of Gap Hamming Distance. Theory of Computing, Vol. 8, 2012, pp. 197-208. www.theoryofcomputing.org

[4] F.A.P. Petitcolas, R.J. Anderson, and M.G. Kuhn, Information Hiding?A Survey, Proc. IEEE, 1999

[5] Al.Jeeva,V.Palanisamy and K. Kanagaram, Comparative Analysis Of Performance Efficiency And Security Measures Of Some Encryption Algorithms, International Journal of Engineering Research and Applications (IJERA) ISSN: 2248-9622 www.ijera.com, Vol. 2, Issue 3, May-Jun 2012, pp.3033-3037.

[6] Cachin, C. An Information-Theoratic Model for Steganography, in Proceedings of the Second International Workshop on Information Hiding, vol. 1525 of Lecture Notes in Computer Science, Springer, 1998, pp. 306-318.

[7] Boris Ryabko and Daniil Ryabko, Constructing perfect steganographic systems, Information and Computation 209 (2011) 1223–1230, © 2011 Elsevier Inc.

[8] Cloudflare. A note about Kerckhoff's Principle. June 19, 2012. Internet:http://blog.cloudflare.com/a-note-about-kerckhoffs-principle

[9] Shirali-Shahreza, M.H. Text Steganography in chat, 3rd IEEE/IFIP International Conference in Central Asia on Internet, (ICI 2007), 2007, pp. 1-5.

[10] Minitab 16. http://www.facebook.com/Minitab [June 6, 2012]

[11] Hamming, Richard W. (1950), Error detecting and error correcting codes, Bell System Technical Journal 29 (2): 147–160, MR 0035935.

[12] Levenshtein VI (1966). Binary codes capable of correcting deletions, insertions, and reversals, Soviet Physics Doklady 10: 707–10.

[13] M.A. Jaro. Probabilistic linkage of large public health data files (disc: P687-689). Statistics in Medicine, 1995, pp. 14:491-498.

[14] The List of Chat Acronyms & Text Message Shorthand, http://www.netlingo.com/acronyms.php

[15] J.Zöllner, H.Federrath, H.Klimant, A.Pfitzmann, R.Piotraschke, A.Westfeld, G.Wicke, and G.Wolf, Modeling the security of steganographic systems, In the Proc. 2nd Workshop on Information Hiding, April 1998, Portland, LNCS 1525, Springer-Verlag, 1998, 345-355.

[16] FIPS PUB 198-1, http://csrc.nist.gov/publications/fips/fips198-1/FIPS-198-1_final.pdf

[17] T Hansen - 2006, US Secure Hash Algorithms (SHA and HMAC-SHA), http://tools.ietf.org/html/rfc4634

[18] Rafat, K. F, Enhanced Text Steganography in SMS, Computer, Control and Communication, 2009, IC4 2009, 2nd International Conference, pp. 1-6

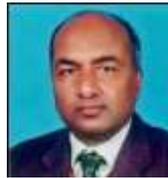

**KHAN FARHAN RAFAT** is a Ph.D. Scholar at International Islamic University, Islamabad – Pakistan. He did MCS from Gomal University, D.I.K. followed by MS in Telecommunication Engineering from UMT, Lahore – Pakistan. As a veteran of information security with almost 24 years of hands-on experience he has worked in varied roles in areas not limited only to programming, evaluation & analysis of Software/Hardware based security modules, and formulating security policies.

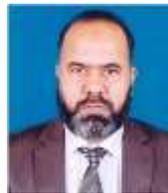

**Professor Dr. Muhammad Sher** is Dean Faculty of Basic and Applied Sciences at International Islamic University, Islamabad – Pakistan. He received B.Sc. degree from Islamia University Bahawalpur and M.Sc. degree from Quaid-e-Azam University, Islamabad, Pakistan. His Ph.D. is from TU Berlin, Germany in Computer Science and Electrical Engineering. His area of research is Next Generation Networks Security. An eminent Scholar who has a number of research publications to his credit.